\documentclass[preprint,showpacs,preprintnumbers,amsmath,amssymb]{revtex4}

\usepackage{graphicx}
\usepackage{dcolumn}
\usepackage{bm}

\begin{document}
\draft
\title {Photoluminescence pressure coefficients of InAs/GaAs quantum dots}
\author {Jun-Wei Luo, Shu-Shen Li, and Jian-Bai Xia}
\address {State Key Laboratory for Superlattices and Microstructures,
Institute of Semiconductors, Chinese Academy of Sciences, P.O. Box
912, Beijing 100083, P.R. China}
\author {Lin-Wang Wang}
\email{lwwang@lbl.gov}
\address {Computational Research Division, Lawrence Berkeley National
Laboratory, Berkeley, CA 94720}
\date{\today}

\begin{abstract}

We have investigated the band-gap pressure coefficients of
self-assembled InAs/GaAs quantum dots by calculating 17 systems
with different quantum dot shape, size, and alloying profile using
atomistic empirical pseudopotential method within the ``strained
linear combination of bulk bands'' approach. Our results confirm
the experimentally observed significant reductions of the band gap
pressure coefficients from the bulk values. We show that the
nonlinear pressure coefficients of the bulk InAs and GaAs are
responsible for these reductions. We also find a rough universal
pressure coefficient versus band gap relationship which agrees
quantitatively with the experimental results. We find linear
relationships between the percentage of electron wavefunction on
the GaAs and the quantum dot band gaps and pressure coefficients.
These linear relationships can be used to get the information of
the electron wavefunctions.

\end{abstract}

\maketitle

Self-assembled InAs quantum dots (QDs) grown on lattice-mismatched
GaAs(100) substrates have been studied extensively in both
experiment and theory in the past 15 years due to their potential
applications and matured synthesise processes ~\cite{Bimberg}.
Depending on synthesise methods and conditions, the quantum dot
can have different size, shape and alloy profile.  A major task of
the research is to study the dependence of the electronic
structure on the size, shape and alloy profile. The electronic
structure includes the electron wavefunctions and their eigen
energies. While there are many experimental ways to probe the
electron eigen energies and their confinement effects [e.g,
photoluminescence (PL) for the exciton energy; the capacitance
charging experiment for Coulomb interaction and the single
particle levels \cite{capacity}], it is much more difficult to
experimentally measure the electronic wavefunctions.
Magnetotunneling spectroscopy~\cite{Vdovin}, low-temperature
scanning tunneling spectroscopy ~\cite{Maltezopoulos}, and
near-field scanning optical microscopes ~\cite{Matsuda} have been
used to probe the electron wavefunctions, but they are not always
successful, and the information about the electron wavefunctions
remain extremely scarce.  Thus any information about the electron
wavefunctions will be extremely useful.

One recently popular experimental approach to study the electronic
structure of a QD is to measure their pressure dependences of the
PL energies. While the PL pressure coefficients (PC) for both bulk
InAs and GaAs are close to 120 $meV/GPa$, it is found
experimentally that the PL pressure coefficients for the quantum
dots are usually much smaller and they can vary significantly from
60 $meV/GPa$ to 100 $meV/GPa$
\cite{Ma,Manjon,Itskevich97,Itskevich98,Itskevich99,Li} depending
on the samples. While Ma {\itshape {et al.}} attributed the main
reason for the much smaller PC to the built-in strain in InAs dots
under nonlinear elasticity theory~\cite{Ma}, Mintairov {\itshape
{et al.}} emphasized the nonuniform In distribution in
QD~\cite{Mintairov}. Thus more quantitative analysis and
understanding are needed here. It is also interesting to find
whether the measured PC of a QD can be used to infer other
properties of the system, e.g., of the electron wavefunctions.

In this letter, via accurate atomistic calculations for the
electron wavefunctions for these quantum dots, we show that the
nonlinear elasticity and the nonlinear band gap pressure
dependence are responsible to the reduction of PC. One problem of
embedded QD study is the lack of reliable experimental information
for the QD size and shape. To overcome this, we have studied 17
different QD systems covering the possible experimental ranges of
QD size and shape. What we find, surprisingly, is a universal
relationship between the QD exciton energy (PL energy) and the
pressure coefficients, which can be compared directly with the
experimental results. Our calculated PC/exciton energy
relationship agrees excellently with the experimental
measurements.  Furthermore, we show that both the QD band gaps and
their PC correlate linearly with the percentage of the electron
wavefunctions on top of the GaAs materials. This is independent of
the QD size, shape, and alloy profile. As a result, these linear
relationships and the corresponding PL and PC experiments can be
used to get the information of the electron wavefunctions.

We will use the empirical pseudopotential method (EPM) \cite{EPM}
to describe the single electron wavefunctions $\psi_i({\bf r})$
of an InAs quantum dot embedded in a GaAs matrix:
\begin{equation}
(-{1\over2} \nabla^2 + V({\bf r})+V_{NL})\psi_i({\bf r})=E_i\psi_i({\bf r}),
\end{equation}
here the total potential $V({\bf r})$ of the system is a direct
sum of the screened atomic empirical pseudopotentals
$\nu_{\alpha}(r)$ of the constituent atoms (type $\alpha$), and
$V_{NL}$ is the nonlocal potential describing the spin-orbit
interaction. The EPM approach has been used to study InAs/GaAs
systems extensively, including quantum dots and alloys. Its
results agree well with experiments \cite{Williamson00}. To study the
various quantum dots in our problem, we need computational
supercells containing up to one million atoms. The wavefunctions
in Eq(1) is expanded by planewave basis. In average, each atom
will have $\sim 50$ planewave basis functions. Thus the Eq(1)
corresponds to a $\sim$ 50 million degree of freedom problem. To
solve Eq(1), we have used the strained linear combination of bulk
band (SLCBB) method \cite{Wang}. In this method, the wavefunction
$\psi_i({\bf r})$ is expanded by bulk Bloch states (which is in
turn expanded by planewaves). Because the bulk Bloch states are
good physical basis functions for the quantum dot states, we can
truncate this basis set (down to $~10,000$) using physical
intuition without introducing significant errors. The errors
caused by the SLCBB method are around 10 meV near the band gap compared with the
exact solution of Eq(1) \cite{Wang_comp}. As a result, this is a much
more accurate method compared to other traditional approaches like
the k.p method, where a few hundred meV error is possible
\cite{Wang_comp}.

To study the pressure effects on the electronic wavefunction, we
first need to study the lattice relaxation under the pressure. We
have used the Keating's valence force field (VFF)
\cite{Pryor,Keating} to described the atomic relaxation. In order
to describe accurately the bulk modulates and their high order
pressure dependence, we have included bond-stretching,
bond-bending, and bond-angle coupling interactions and high order
bond-stretching terms ~\cite{Williamson00}. Table I lists the VFF
bulk modulates and their pressure dependence. They agree well with
the experiments. To be able to describe accurately the nonlinear
lattice relaxation is important because there is a $\sim7.2\%$
lattice mismatch between bulk InAs and GaAs. For a quantum dot
system, InAs is under compressive stress and GaAs is under tensile
stress. They will behave differently under additional external
pressure because of the nonlinear lattice relaxation.

After the atomic relaxation is described accurately by the VFF
model, the pressure dependence of the bulk band structures for
GaAs and InAs is described by the EPM Hamiltonian. Here, an
explicit local strain dependence of $\nu_{\alpha}(r)$ is used to
describe accurately the deformation potentials of the band
energies \cite{Williamson00}. Thus the fitting of
$\nu_{\alpha}(r)$ not only provide an accurate band structure at
zero pressure, it also provides accurate high order pressure
dependence of the band energies. Fig.1 shows the calculated band
energy pressure dependence for bulk InAs and GaAs. The calculated
band gap pressure coefficients for InAs and GaAs are 117 and 103
meV/GPa respectively, they agree well with the experimental values
of 114 and 106(4) meV/GPa~\cite{LB}.

We next use the above VFF and EPM Hamiltonians to calculate
various embedded quantum dots under different pressures. A large
variety of QD shapes have been reported and studied for the
InAs/GaAs system by various groups, for example pyramidal quantum
dot (PQD) with side facets oriented along \{101\}, \{113\}, or
\{105\}~\cite{Kim} or truncated pyramidal quantum dot
(TPQD)~\cite{Bruls,Liu}. Inside the QD, various In/Ga profiles
have been speculated, for example an inverted-triangle shape
In-rich core ~\cite{Liu} or a growth direction linearly increasing
In concentration ~\cite{Bruls}. To cover the whole spectrum of
possible shapes and alloy profiles, we have used three sets of
QDs: pure pyramidal QDs with \{101\}, or \{113\}, or \{105\}
facets; pure truncated pyramidal QDs with different height/base
ratios; and linearly increasing In concentration alloy profile
QDs. Besides the shapes and alloy profiles, different sizes of the
same shape QD are used. In total we have calculated 17 different
quantum dots, their sizes, shapes and alloy profiles are described
in Table II.

The above described InAs quantum dots are embedded in a pure GaAs
matrix. A supercell box is used to contain the quantum dot. A
periodic boundary condition is used for the supercell box. To
remove the possible dot-dot electronic and elastic interactions,
sufficient GaAs barrier is used. As a result, a supercell can
contain upto one million atoms. The atomic positions within the
supercell are then relaxed by minimizing the strain energy of the
VFF Hamiltonian. To create a pressure, the overall size of the
supercell is changed, and the pressure is calculated from the
local GaAs strains away from the quantum dot. After the atomic
positions are relaxed, the electron and hole eigenstates and eigen
energies of Eq(1) are solved using the SLCBB method.

We typically calculate 5 pressure values from 0 to 2 GPa for each
quantum dot. Using these five points, the band gap of the quantum
dot is fitted as $E_g(P)=E_g(0)+a_1 P + a_2 P^2$. Then the linear
pressure coefficients (PC) of the band gap is read out from $a_1$.
In consistent with the experiment, we find this PC is in the range
of 60-110 meV/GPa, much smaller than the bulk InAs and GaAs PC. We
then plot all the calculated PC as a function of the QD zero
pressure exciton energy $E_0(0)$ (which is the band gap minus the
electron hole interaction),the result is shown in Fig.2.
Surprisingly, despite all the different shapes and sizes for the
17 QDs we studied, we find a rough universal linear relationship
between the PC and the exciton energy. This provides a convenient
way to compare with the experiment, without the need to know the
QD size and shape which are not available from the experiment. The
theory and experiment comparison is shown in Fig.2. The agreement
is excellent considering all the possible uncertainties involved.
We see that, indeed, the QD pressure coefficients are much smaller
than the bulk values of both InAs and GaAs, and they decrease with
the exciton energy.

To understand the variation of the PC, and its dependence on the
QD, we can perform a simple analysis. We will concentrate on the
conduction band minimum (CBM) state since most of the band gap
pressure coefficient comes from the conduction band \cite{Wei}.
For a simple approximation, we can express the energy $E_{CBM}$ of
the CBM eigenstate $\psi_{CBM}({\bf r})$ as a sum of an effective
mass like potential energy and a kinetic energy $E_k$, and the
potential energy can be approximated by a weighted sum of the
local conduction band energy:

\begin{equation}
 E_{CBM} \approx \int |\psi_{CBM}({\bf r})|^2 E_c({\bf r})  d^3r + E_k ,
\end{equation}
here the $E_c({\bf r})$ is the bulk conduction band energy for the
given local strain at ${\bf r}$ and the local constituent material
(either GaAs or InAs). Note that, in practice, the space integral
of Eq(2) is replaced by a sum over the atom $\sum_{at} W_{at}
E_c(at)$, where the local strain for an atom is calculated from
the atom's nearest neighbor atomic positions, and $W_{at}$ denotes
the weight of $|\psi_{CBM}({\bf r})|^2$ at that atom $"at"$. We
have plotted $E_{CBM}$ as a function of $\sum_{at} W_{at} E_c(at)$
in Fig.3(a). We see that all the calculated QDs fall into a nice
curve. The difference between this curve and the dashed line (the
potential energy line) is the kinetic energy $E_k$.

Now, we analysis the pressure coefficients of $E_{CBM}$ using
Eq(2). If we ignore the pressure dependences of the kinetic energy
and the weight function $W_{at}$, we can have an approximated
relationship:
\begin{equation}
 E'_{CBM} \approx \int |\psi_{CBM}({\bf r})|^2 E'_c({\bf r})  d^3r ,
\end{equation}
here the prime indicate the derivation with pressure. Despite all
the approximations, the left and right hand side of Eq(3) do form
a nice linear relationship, as shown in Fig.3(b). The slope of the
line in Fig.3(b) is not 1, but 1.25, indicating the right hand
side of Eq(3) account only for about $80\%$ of the left hand side.
This situation can be compared with the case of free standing
colloidal quantum dots \cite{JLi}, where the change of PC in a QD
can be traced back completely from their bulk origin. Our current
embedded QD is much more complicated due to the internal strain
effects between InAs and GaAs, we find such accurate analysis is
impossible here.

Despite of not accounting $100\%$ of the left hand side in Eq(3),
the physical meaning of the right hand side of Eq(3) is clear and
useful [especially when it is written as $E'_{CBM} \approx
\sum_{at} W_{at} E'_c(at)$]: the PC of the quantum dot state is a
wavefunction weighted sum of the local PC at all the atoms. The
$E'_c(at)$ depends on the local strain of this atom as illustrated
in Fig.1. This can be used to understand why the QD PC is in
general less than the bulk InAs and GaAs results. Because InAs in
the QD is always under compressive strain, due to the nonlinear PC
as shown in Fig.1, the $E'_c$ in the InAs region is significantly
smaller than its bulk value of 130 meV/GPa. On the other hand,
GaAs is under tensile strain, which will increase $E'_c$. However,
because the magnitude of the GaAs strain is in general smaller
than the InAs strain, and because most wavefunction is localized
in the InAs region, the averaged PC is then smaller than the bulk
InAs and GaAs PCs. Thus we see that the nonlinear bulk PC is
responsible for the reduction of QD PC compared to bulk values, in
consistent with the explanation provided by Ma {\it et
al}~\cite{Ma}.

Guided by Eqs(2),(3), we now try to find some simple relationships
between the experimentally easily observable quantities (band gap
and pressure coefficients) and the wavefunction properties. In
Eq(3), if we represent $E_c'(at)$ by just two values, one for
InAs, one for GaAs, then $E'_{CBM}$ of Eq(3) becomes an linear
function of $x=\sum_{at\in GaAs} W_{at}/\sum_{at\in all} W_{at}$
(i.e, the percentage of the wavefunction on GaAs). This hypothesis
is tested in Fig.4(a), where we have plotted the pressure
coefficients of the exciton energy (not just the CBM energy), so
the connection with experiment is more straightforward. We see
that $E'_0$ and $x$ form a very nice straight line. This can be
very useful, since a measured $E'_0$ value will give us the $x$,
which is a property of the wavefunction that cannot be measured
easily by other means. The same relationship can be plotted
between the exciton energy itself $E_0$ and the $x$, as in
Fig.4(b). They also form a rough linear relationship although with
larger scatters. The linear relationships in Fig.4(a) and (b), in
turn, explain why we have a rough universal relationship between
$E_0'$ and $E_0$ in Fig.2. This is because both $E_0'$ and
$E_0$ are linearly correlated with x.

In summary, using accurate and reliable empirical pseudopotential
methods and the SLCBB calculations, we have studied InAs/GaAs
quantum dot PL pressure coefficients. We investigated 17 different
quantum dots covering the ranges of experimental QD size, shape
and alloy profile. We found a universal PC and exciton energy
relationship, which agrees excellently with the experimental
results. We also find linear relationships between the
wavefunction percentage on GaAs and the PL pressure coefficient
and PL energy. These linear relationships can be used to probe the
properties of the electron wavefunctions.

We would like to acknowledge G.H. Li and B.S. Ma for helpful
discussions. Part of the CPU-time of this work was supplied by
Supercomputing Center, CNIC, CAS. This work was supported by the
National Natural Science Foundation of China and the special funds
for Major State Basic Research Project No. G2001CB309500 of China.
The work by L.W. Wang is also funded by U.S. Department of Energy under
Contract No. DE-AC03-76SF00098.


\newpage

\begin{table}
\caption{\label{tab:table1} The VFF bulk modulates and their first
and second order pressure coefficients. The bulk modulates are in
the unit of $10GPa$, the $dB/dP$ has a unit 1, and $d^2B/dP^2$ is
in the unit of $GPa^{-1}$.}
\begin{ruledtabular}
\begin{tabular}{ccccc}
& \multicolumn{2}{c}{GaAs} & \multicolumn{2}{c}{InAs} \\
\cline{2-3}\cline{4-5} Property &Fitted
&Expt.\tablenote{Reference ~\cite{LB}} &Fitted & Expt.$^a$\\
\hline
$C_{11}$ &12.11 &12.11(4) &8.328 &8.329 \\
$C_{12}$ &5.50  &5.48(17) &4.553 &4.526 \\
$C_{44}$ &6.04  &6.04(2)  &3.803 &3.959 \\
$B$      &7.70  &7.54     &5.811 &5.794\\
$dB/dP$  &5.01  &4.49     &5.329 &4.787\\
$d^2B/dP^2$  &-0.111  &---     &-0.144 &---\\
\end{tabular}
\end{ruledtabular}
\end{table}

\begin{table}
\caption{\label{tab:table2} The 17 calculated quantum dots}
\begin{ruledtabular}
\begin{tabular}{lcccccccc}
    & \multicolumn{7}{c}{ Pure InAs pyramidal quantum dots (PQDs)}   \\
    & 1 & 2 & 3 & 4 & 5 & 6 & 7  \\
facet  &  \{101\} & \{101\} & \{113\} & \{113\} & \{105\} &\{105\} &\{105\}\\
base size(nm)  & 6 & 11.3 & 6 & 11.3 & 11.3 & 15 & 20 \\
\hline
    & \multicolumn{8}{c}{ Pure InAs truncated pyramidal quantum dots (TPQDs)}   \\
    & 8 & 9 & 10 & 11 & 12 & 13 & 14 &15  \\
facet  &\{101\} &\{101\} &\{101\} &\{101\} &\{101\} &\{101\} &\{101\} &\{101\}\\
base size(nm)  & 6 & 6 & 6 & 11.3 & 11.3 & 11.3 & 11.3 &11.3 \\
height/base & 2/3 & 1/2 & 1/4 & 2/3 & 1/2 & 1/4 & 1/5 & 1/10 \\
\hline
    & \multicolumn{6}{c}{Alloy pyramidal quantum dots}   \\
    & \multicolumn{3}{c}{16} &\multicolumn{3}{c}{17}\\
facet  & \multicolumn{3}{c} {\{101\}} & \multicolumn{3}{c}{\{101\}} \\
base size(nm)  & \multicolumn{3}{c}{11.3} & \multicolumn{3}{c}{11.3}   \\
alloy profile &\multicolumn{3}{c}{bottom 40\%Ga, tip 0\%Ga}
&\multicolumn{3}{c}{bottom 50\%Ga, tip 0\%Ga} \\
\end{tabular}
\end{ruledtabular}
\end{table}

\begin{figure}
\begin{center}
\includegraphics[width=0.6\textwidth]{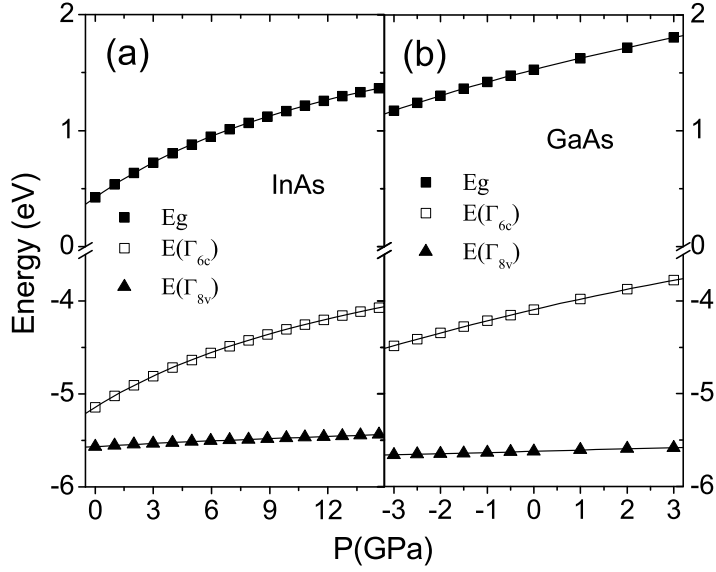}
\end{center}
\caption{The band-edge energies ($E(\Gamma{_{6c}})$, and
$E(\Gamma{_{8v}})$) of (a) bulk InAs and (b) GaAs and their direct
band gap $E_g(\Gamma{_{8v}}-\Gamma{_{6c}})$ under hydrostatic
pressure.} \label{fig1}
\end{figure}

\begin{figure}
\includegraphics[width=0.6\textwidth]{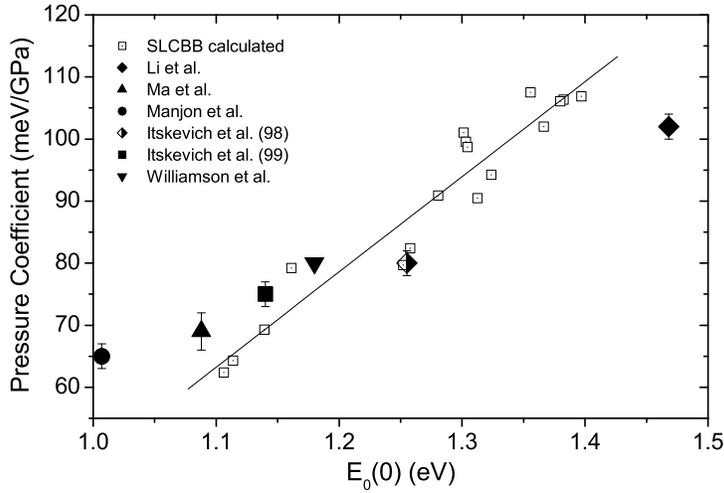}
\caption{\label{fig2} The PL pressure coefficient ($E'_0$) versus
$E_0(0)$ (PL energy) and comparison with experiments. The $E_0(0)$ is
the zero pressure exciton energy which equals the band gap minus
the electron hole Coulomb interaction. The experimental results
are: Li {\itshape {et al.}}~\cite{Li}, Ma {\itshape {et
al.}}~\cite{Ma}, Manjon {\itshape {et al.}}~\cite{Manjon}, and
Itskevich {\itshape {et al.}}~\cite{Itskevich98,Itskevich99}.  We
also included one previously calculated result from Williamson
{\itshape {et al.}}~\cite{Williamson98}.}
\end{figure}

\begin{figure}
\includegraphics[width=0.6\textwidth]{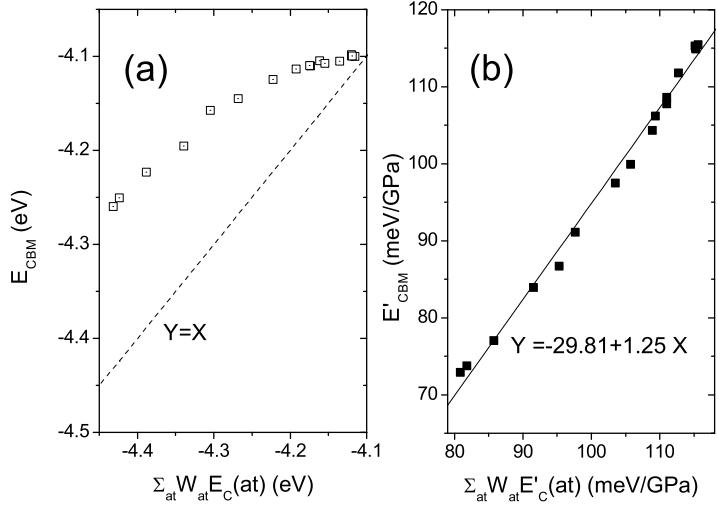}
\caption{\label{fig3} (a) The $E_{CBM}$ as a function of
$\sum_{at} W_{at} E_c(at)$; (b)The $E'_{CBM}$ as a function of
$\sum_{at} W_{at} E'_c(at)$. }
\end{figure}

\begin{figure}
\includegraphics[width=0.6\textwidth]{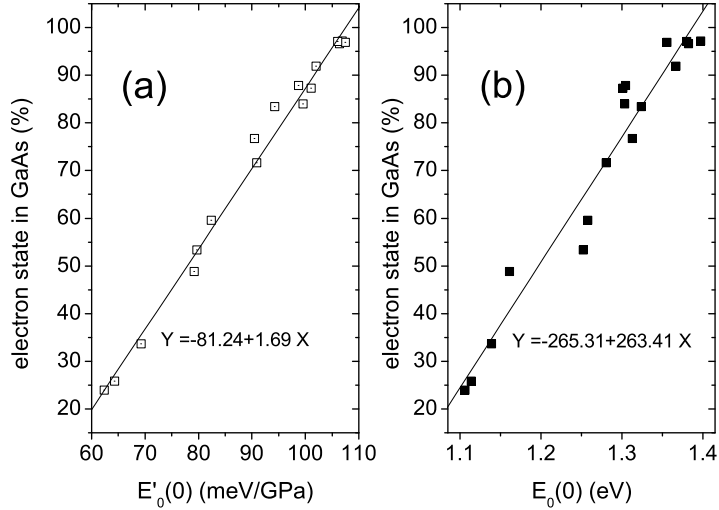}
\caption{\label{fig4} (a)The relationship between $E'_0(0)$ and x
(the percentage of the electron state in GaAs); (b) The
relationship between $E_0(0)$ and x.}
\end{figure}

\end{document}